\documentclass[lettersize,journal]{IEEEtran}
\usepackage{amsmath,amsfonts,amssymb}
\usepackage{bm}
\usepackage{cite}
\usepackage{graphicx}
\usepackage{subfigure}

\title{Misspecified Cramér--Rao Bound for AoA Estimation at a ULA under a Spoofing Attack}
\author{Sotiris Skaperas,~\IEEEmembership{Member,~IEEE}, Arsenia Chorti,~\IEEEmembership{Senior Member,~IEEE}%

}

\begin{document}
\maketitle

\begin{abstract}
A framework is presented for analyzing the impact of active attacks to location-based physical layer authentication  (PLA) using the machinery  of  misspecified Cramér--Rao bound (MCRB). In this work, we focus on the MCRB in the angle-of-arrival (AoA) based authentication of a single antenna user when the verifier posseses an $M$ antenna element uniform linear array (ULA), assuming deterministic pilot signals; in our system model the presence of a spoofing adversary with an arbitrary number $L$ of antenna elements is assumed. We obtain a closed-form expression for the MCRB and demonstrate that the attack introduces in it a penalty term compared to the classic CRB, which does not depend on the signal-to-noise ratio (SNR) but on the adversary's location, the array geometry and the attacker precoding vector.
\end{abstract}


\section{Introduction}

Angle-of-arrival (AoA) estimation in digital array systems is fundamental to localization and emerging sensing. Recently, it has been shown that the AoA can also serve as a robust feature against spoofing (impersonation) attacks in location-based physical layer authentication (PLA) \cite{hoang2024physical}. For example, in \cite{Pham_Globecom23},\cite{Pham_arxiv}, it was shown that an irreducible error floor is introduced in the mean square error between the legitimate and adversarial observations. 

In the present work, we extend this analysis to derive the fundamental limit in the AoA estimation under the adversarial model in the  case of a verifier using a uniform linear array (ULA) with $M$ antenna elements. The motivation behind this analysis lies in the fact that when the  observations are not generated from the estimator's assumed model, due to an adversary attempting an impersonation attack, classical performance bounds such as the Cramér--Rao bound (CRB) fail to provide a reliable measure of achievable accuracy, as they ignore the underlying model mismatch \cite{fortunati2016misspecified}. In this context, misspecified Cramér–Rao bound (MCRB) has emerged as a vital framework to characterize estimator performance under model mismatch in communication and sensing applications \cite{Varughese2025MCRB}, \cite{abed2021misspecified}. While MCRB has been applied to geometric mismatches in reconfigurable intelligent surfaces (RIS)-aided localization \cite{Zheng2023MCRB} and channel estimation, its application to adversarial spoofing in ULAs remains is largely unexplored, an exception is given in \cite{Usman2025}.

Motivated by the above, we derive the MCRB for AoA estimation under a spoofing attack. The main contributions of this paper are as follows:
\begin{itemize}
    \item We derive the MCRB for AoA estimation in the presence of an impersonation attack, providing a rigorous quantification of estimation limits under model mismatch.
    \item We provide a closed-form expression for the MCRB, isolating a specific "mismatch penalty" that characterizes the cost of the adversarial presence.
    \item Our analysis demonstrates that the MCRB is always greater than or equal to the CRB due to a penalty term that does not depend on the signal-to-noise ratio (SNR) but rather on the adversary's actual location (angular separation), its precoding vector and the uniform linear array (ULA) geometry ($M^2$ scaling as opposed to $M^3$ scaling for the standard term).
\end{itemize}    

The remainder of the paper is organized as follows. Section~II introduces the assumed and true signal models under spoofing. Section~III derives a closed-form expression of the MCRB for AoA estimation at a ULA, along with a discussion of its main properties. Section~IV presents numerical and Monte--Carlo results and Section~V concludes the paper.

\section{System Model}

We consider a ULA with $M$ elements, inter-element spacing $d$ and operating wavelength $\lambda$, we further denote the wavenumber by $\kappa\triangleq 2\pi d/\lambda$. Without loss of generality, we assume unitary (deterministic) pilot signals are exchanged in a  far-field narrowband channel scenario.

\subsection{Assumed (Estimation) Model}

Given the verifier observes a single snapshot generated by a single-path Gaussian noise model, the received signal is, 
\begin{equation}
    \mathbf{x} = \mathbf{a}(\theta) + \mathbf{n},
    \label{eq:assumed}
\end{equation}

where \(\mathbf{x}\in\mathbb{C}^M\) is the observed snapshot, \(\theta\in[-\pi/2,\pi/2]\) is the AoA, and 
\(\mathbf{n}\sim\mathcal{CN}(\mathbf{0},\sigma^{2}\mathbf{I}_M)\) is spatially white circularly symmetric
complex Gaussian noise with variance $\sigma^{2}$ per antenna. The ULA steering vector is given by,
\begin{equation}
    \mathbf{a}(\theta)=\left[1,\,e^{-j\kappa\sin\theta},\,\ldots,\,e^{-j\kappa(M-1)\sin\theta}\right]^{T}.
\end{equation}
In the following we use dot notation for derivatives with respect to \(\theta\), e.g., \(\dot{\mathbf a}(\theta)\triangleq\partial\mathbf a(\theta)/\partial\theta\).

\subsection{True (spoofed) Model}

The actual (spoofed) received signal is denoted by,
\begin{equation}
    \hat{\mathbf{x}} = \mathbf{s} + \mathbf{n},
    \label{eq:true}
\end{equation}
where the noise vector \(\mathbf{n}\sim\mathcal{CN}(\mathbf{0},\sigma^{2}\mathbf{I}_M)\) has the same statistics as in the assumed model and the true (spoofed) mean signal is modeled as in \cite{Pham_arxiv},
\begin{equation}
    \mathbf{s} \triangleq \sum_{\ell=1}^{L} q_{\ell}\,\mathbf{a}(\hat{\theta}_{\ell}),
\end{equation}
corresponding to an attacker with $L$ distributed antennas,
using complex precoding coefficients \(q_{\ell}\in\mathbb{C}\) and transmitting from the true angles \(\hat{\theta}_{\ell}\in[-\pi/2,\pi/2]\).

We define the mismatch vector
\begin{equation}
    \boldsymbol{\Delta}(\theta)\triangleq \mathbf{s}-\mathbf{a}(\theta).
    \label{eq:delta_def}
\end{equation}

\section{Misspecified Cram\'er Rao Bound}

We start from the likelihood function of the assumed model in \eqref{eq:assumed}, which under the Gaussian  distribution of \(\mathbf{x}\) conditioned on \(\theta\) is given by
\begin{equation}
    p(\mathbf{x};\theta)
    = \frac{1}{(\pi\sigma^{2})^{M}}
      \exp\!\left(-\frac{1}{\sigma^{2}}
      \|\mathbf{x}-\mathbf{a}(\theta)\|^{2}\right),
    \label{eq:likelihood}
\end{equation}
and the corresponding log-likelihood is
\begin{equation}
    \ln p(\mathbf{x};\theta)
    = -M\ln(\pi\sigma^{2})
      - \frac{1}{\sigma^{2}}\|\mathbf{x}-\mathbf{a}(\theta)\|^{2}.
\end{equation}

We differentiate the log-likelihood with respect to \(\theta\) to obtain the score function:
\begin{align}
    u(\mathbf{x};\theta)
    &\triangleq \frac{\partial}{\partial\theta}\ln p(\mathbf{x};\theta)
    = -\frac{1}{\sigma^{2}}\frac{\partial}{\partial\theta}\left(\mathbf{x}-\mathbf{a}(\theta)\right)^{H}(\mathbf{x}-\mathbf{a}(\theta)) \nonumber\\
    &= -\frac{1}{\sigma^{2}}\left[
        -\dot{\mathbf{a}}(\theta)^{H}(\mathbf{x}-\mathbf{a}(\theta))
        -(\mathbf{x}-\mathbf{a}(\theta))^{H}\dot{\mathbf{a}}(\theta)
    \right] \nonumber\\
    &= \frac{2}{\sigma^{2}}\Re\!\left\{\dot{\mathbf{a}}(\theta)^{H}(\mathbf{x}-\mathbf{a}(\theta))\right\}.
    \label{eq:score}
\end{align}
where, since \(\theta\) is a real-valued parameter, eq. \eqref{eq:score} uses the identity \(z+z^{*}=2\Re\{z\}\) for a complex scalar \(z\).

\subsection{Fisher Information under the Assumed Model}

Under the assumed model in \eqref{eq:assumed}, the observed information, defined as the negative second derivative of the log-likelihood, is given by
\begin{align}
    -\frac{\partial^{2}}{\partial\theta^{2}}\ln p(\mathbf{x};\theta)
    &= \frac{2}{\sigma^{2}}\Re\!\left\{ \dot{\mathbf{a}}(\theta)^{H}\dot{\mathbf{a}}(\theta) - \ddot{\mathbf{a}}(\theta)^{H}(\mathbf{x}-\mathbf{a}(\theta)) \right\},
\end{align}
where \(\ddot{\mathbf a}(\theta)\triangleq\partial^2\mathbf a(\theta)/\partial\theta^2\).
Evaluating the expectation with respect to the assumed model \(p(\mathbf{x};\theta)\), so that \(\mathbb{E}[\mathbf{x}\mid\theta]=\mathbf{a}(\theta)\), the second term vanishes and the Fisher information reduces to
\begin{equation}
    J(\theta) \triangleq \mathbb{E}_{\mathrm{assumed}}\left[-\frac{\partial^{2}}{\partial\theta^{2}}\ln p(\mathbf{x};\theta)\right]
    = \frac{2}{\sigma^{2}}\|\dot{\mathbf{a}}(\theta)\|^{2}.
    \label{eq:J_def}
\end{equation}
We denote by
\begin{equation}
    \Gamma(\theta)\triangleq \|\dot{\mathbf{a}}(\theta)\|^{2}.
    \label{eq:Gamma_def}
\end{equation}

In the case of an ULA the explicit expression for \(\Gamma(\theta)\) is derived as
\begin{equation}
    \Gamma(\theta) = \kappa^{2}\cos^{2}\theta \sum_{m=0}^{M-1} m^{2}
    = \kappa^{2}\cos^{2}\theta  \frac{(M-1)M(2M-1)}{6}.
    \label{eq:Gamma_ULA}
\end{equation}

\subsection{Second Moment of the Score under the True Model}

Under the model mismatch, the second order moment of the score needs to be evaluated
under the true distribution of \(\hat{\mathbf{x}}\), i.e.,
\begin{equation}
    K(\theta)\triangleq \mathbb{E}_{\mathrm{true}}\left[u(\hat{\mathbf{x}};\theta)^{2}\right].
    \label{eq:K_def}
\end{equation}

The mean of the score under the true model is given by
\begin{align}
    \mathbb{E}_{\mathrm{true}}[u(\hat{\mathbf{x}};\theta)]
    &= \frac{2}{\sigma^{2}}\Re\!\left\{\dot{\mathbf{a}}(\theta)^{H}(\mathbb{E}_{\mathrm{true}}[\hat{\mathbf{x}}]-\mathbf{a}(\theta))\right\} \nonumber\\
    &= \frac{2}{\sigma^{2}}\Re\!\left\{\dot{\mathbf{a}}(\theta)^{H}\boldsymbol{\Delta}(\theta)\right\},
    \label{eq:score_mean}
\end{align}
where \(\boldsymbol{\Delta}(\theta)\) is defined in \eqref{eq:delta_def}.
Using the identity
\(\mathbb{E}[u^{2}]=\operatorname{Var}[u]+(\mathbb{E}[u])^{2}\),
we next compute the variance of the score under the true model.

We note that the randomness in
\(u(\hat{\mathbf{x}};\theta)\) arises solely from the noise component in
\(\hat{\mathbf{x}}=\mathbf{s}+\mathbf{n}\), and that the noise vector
\(\mathbf{n}\sim\mathcal{CN}(\mathbf{0},\sigma^{2}\mathbf{I}_M)\) has the same
covariance under both the assumed and the true models. Consequently, we have
\begin{align}
    \operatorname{Var}_{\mathrm{true}}[u(\hat{\mathbf{x}};\theta)]
    &= \frac{4}{\sigma^{4}}
       \operatorname{Var}\!\left(
       \Re\{\dot{\mathbf{a}}(\theta)^{H}\mathbf{n}\}\right) \nonumber\\
    &= \frac{4}{\sigma^{4}}\cdot\frac{1}{2}\,
       \mathbb{E}\!\left\{
       \left|\dot{\mathbf{a}}(\theta)^{H}\mathbf{n}\right|^{2}\right\}
       \nonumber\\
    &= \frac{2}{\sigma^{4}}\cdot\sigma^{2}
       \|\dot{\mathbf{a}}(\theta)\|^{2} \nonumber\\
    &= \frac{2}{\sigma^{2}}\Gamma(\theta),
    \label{eq:score_var}
\end{align}
where we used the fact that for a complex Gaussian random variable
\(z=\dot{\mathbf{a}}^{H}\mathbf{n}\), \(\operatorname{Var}(\Re\{z\})=\tfrac{1}{2}\mathbb{E}[|z|^{2}]\), and \(\mathbb{E}[|z|^{2}]=\sigma^{2}\|\dot{\mathbf{a}}(\theta)\|^{2}\).

Combining \eqref{eq:score_mean} and \eqref{eq:score_var} yields the second moment
\begin{align}
    K(\theta) &= \operatorname{Var}_{\mathrm{true}}[u(\hat{\mathbf{x}};\theta)] + \big(\mathbb{E}_{\mathrm{true}}[u(\hat{\mathbf{x}};\theta)]\big)^{2} \nonumber\\
    &= \frac{2}{\sigma^{2}}\Gamma(\theta) + \left(\frac{2}{\sigma^{2}}\Re\{\dot{\mathbf a}(\theta)^{H}\boldsymbol{\Delta}(\theta)\}\right)^{2}.
    \label{eq:K_final}
\end{align}

Define the scalar
\begin{equation}
    \eta(\theta)\triangleq
    \Re\{\dot{\mathbf a}(\theta)^{H}\boldsymbol{\Delta}(\theta)\},
    \label{eq:eta_def_repeat}
\end{equation}
which quantifies the mismatch induced by the spoofed mean signal. Then, \eqref{eq:K_final} can be written as
\begin{equation}
    K(\theta)=\frac{2}{\sigma^{2}}\Gamma(\theta)
    + \left(\frac{2}{\sigma^{2}}\eta(\theta)\right)^{2}.
\end{equation}
where
\begin{equation}
    r_{\ell}\triangleq e^{j\kappa(\sin\hat{\theta}_{\ell}-\sin\theta)}.
\end{equation}
For \(r_{\ell}\neq 1\), the finite weighted geometric sum
\begin{equation}
    S_M(r_{\ell})\triangleq \sum_{m=0}^{M-1} m r_{\ell}^{m}
    = \frac{r_{\ell}\big(1 - M r_{\ell}^{M-1} + (M-1) r_{\ell}^{M}\big)}{(1-r_{\ell})^{2}}
\end{equation}
arises from the inner product \(\dot{\mathbf a}(\theta)^{H}\mathbf a(\hat{\theta}_{\ell})\).
For \(r_{\ell}=1\), corresponding to \(\hat{\theta}_{\ell}=\theta\), the limit yields
\(S_M(1)=M(M-1)/2\).

\subsection{MCRB}

The MCRB for the scalar parameter \(\theta\) is given by the following form
\begin{equation}
    \mathrm{MCRB}(\theta)=J(\theta)^{-1}K(\theta)J(\theta)^{-1}.
    \label{eq:mcrb_gen}
\end{equation}
where \(J(\theta)\) denotes the Fisher information of the assumed model, while \(K(\theta)\) is the second-order moment of the score evaluated under the true (spoofed) model.

Using \(J(\theta)=\frac{2}{\sigma^{2}}\Gamma(\theta)\) from
\eqref{eq:J_def} and the expression for \(K(\theta)\) in
\eqref{eq:K_final}, the MCRB can be written in the decomposed form
\begin{equation}
    \mathrm{MCRB}(\theta)
    = \underbrace{\frac{\sigma^{2}}{2\Gamma(\theta)}}_{\text{CRB}}
    + \underbrace{\frac{\eta(\theta)^{2}}{\Gamma(\theta)^{2}}}_{\text{mismatch penalty}}.
    \label{eq:mcrb_result}
\end{equation}

The first term in \eqref{eq:mcrb_result} matches with the classical CRB obtained, representing the bound on the variance under the assumed model. The second term is a non-negative penalty that reflects the mismatch between the assumed and true signal models (due to spoofing) and vanishes when the mismatch is absent, i.e.,\(\boldsymbol{\Delta}(\theta)=\mathbf{0}\).

Substituting the explicit ULA expression for \(\Gamma(\theta)\) yields the
classical CRB
\begin{equation}
    \mathrm{CRB}(\theta)
    = \frac{3\sigma^{2}}{
      \kappa^{2}\cos^{2}\theta\,(M-1)M(2M-1)}.
\end{equation}
The mismatch-induced term depends solely on the spoofer's configuration through \(\eta(\theta)\). Substituting \(\Gamma(\theta)\) and \(\eta(\theta)\) yields the closed-form solution
\begin{align}
\mathrm{MCRB}(\theta)
&=
\frac{3\sigma^{2}}{\kappa^{2}\cos^{2}\theta\,(M-1)M(2M-1)}
\nonumber\\ 
&+
\frac{
\left(
\Im\!\left\{\sum_{\ell=1}^{L} q_{\ell}\,S_{M}(r_{\ell})\right\}
\right)^{2}
}{
\kappa^{2}\cos^{2}\theta
\left(\frac{(M-1)M(2M-1)}{6}\right)^{2}
}.
\label{eq:mcrb_closed}
\end{align}

\subsection{Properties of the derived MCRB}
The decomposed expression of the MCRB in \eqref{eq:mcrb_result} reveals the following key remarks:

\begin{itemize}
\item \textit{MCRB is always greater than or equal to CRB} since $\mathrm{MCRB}(\theta)-\mathrm{CRB}(\theta)=\frac{\eta(\theta)^2}{\Gamma(\theta)^2}\ge 0$, with equality if and only if $\eta(\theta)=0$. This condition is satisfied in the absence of model mismatch (\(\boldsymbol{\Delta}(\theta)=\mathbf 0\)), but may also occur under special geometric (L, M and $\kappa$) or precoding configurations (\(q_{\ell}\) and \(\hat{\theta}_{\ell}\)) for which the mismatch vector is orthogonal (in the real-part sense) to \(\dot{\mathbf a}(\theta)\).

\item \textit{Dependence on attacker configuration (not SNR).} The mismatch penalty depends exclusively on the attacker configuration through $\{\hat{\theta}_\ell,q_\ell\}$ and the array geometry via $M$ and $\kappa$, as captured by $\eta(\theta)$ in \eqref{eq:eta_final}, and is independent of $\sigma^2$. In particular, $\eta(\theta)=0$ occurs only under specific angular and phase coordination conditions.

\item \textit{Perfect alignment exemption.} If the adversary aligns exactly with the assumed steering direction (\(\hat{\theta}_{\ell}=\theta, \ell=L=1\)), then \(r_{\ell}=1\) and the corresponding \(S_M(1)=M(M-1)/2\) yields a purely imaginary \(\dot{\mathbf a}(\theta)^{H}\mathbf a(\theta)\) contribution; such aligned terms do not contribute to \(\eta(\theta)\) (their real part is zero). Thus, an attacker on the same direction as the legitimate user does not introduce any model mismatch and cannot be potentially identified, based solely on AoA.

\item \textit{Impact of array size M.}
For a ULA, $\Gamma(\theta)=\Theta(M^3)$ due to the quadratic growth of $\sum m^2$, while $(|S_M(r)|\le \sum_{m=0}^{M-1} m=\Theta(M^2)$. For bounded $\{q_\ell\}$ and fixed $L$, the mismatch-induced term scales as $O(M^{-2})$, whereas the classical CRB decays faster as $O(M^{-3})$. Thus, although increasing $M$ improves estimation accuracy, the relative importance of the mismatch penalty increases at large array sizes.

\item \textit{}
\textit{Impact of attacker components size L.}
Increasing the number of attacker components can increase the mismatch-induced error floor through coherent accumulation of their spatial signatures.
Nevertheless, since the Fisher information grows faster with the array size than the mismatch term, the worst-case penalty scales at most as $O(L^2/M^2)$, or $O(L/M^2)$ under fixed total attacker power, and therefore diminishes for sufficiently large arrays.

\end{itemize}

\section{Numerical Results}

In this section, we evaluate the derived MCRB and illustrate its behavior under spoofing attacks through numerical and  simulations results. We assume a ULA with half-wavelength spacing $d=\lambda/2$ (i.e., $\kappa=\pi$) and AoA for the ligitimate transmitter \(\theta=10^\circ\), in all assessments. Unless otherwise stated, we consider $M=16$ antenna elements, $L=1$ spoofing components and an SNR$\in[0,50]$~dB.

\begin{figure}[t]
    \centering
    \includegraphics[width=\linewidth]{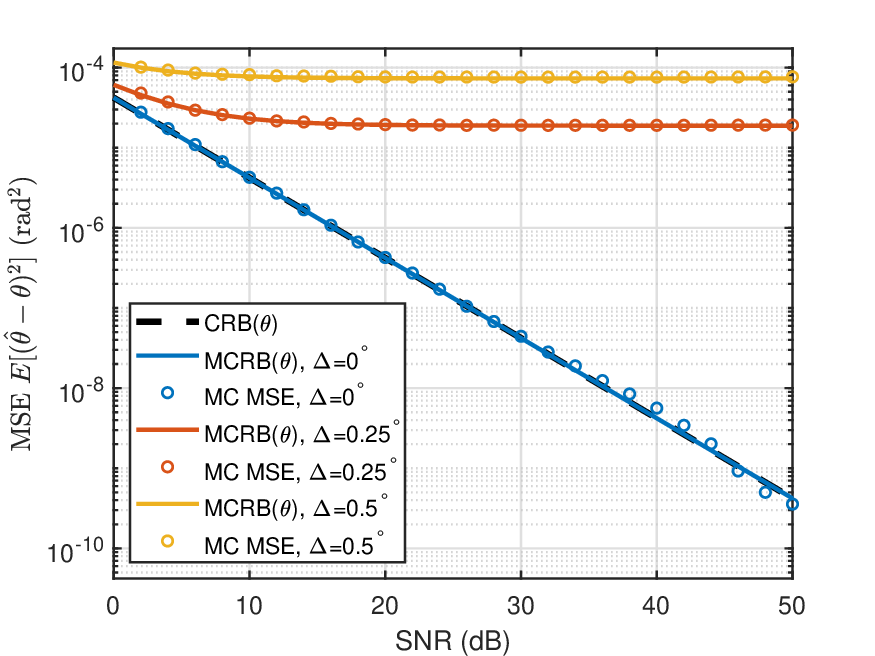}
    \caption{Monte--Carlo MSE versus SNR for AoA estimation under spoofing with $\Delta=\{0^{\circ},0.25^{\circ},0.5^{\circ}\}$ and $M=16$.}
    \label{fig:mc_validation}
\end{figure}

Initially, in Fig.~\ref{fig:mc_validation} we compare the MSE $\mathbb{E}[(\hat\theta-\theta)^2]$ of the AoA estimator under various attacker angular separations $\Delta=\hat{\theta}-\theta\in\{0^\circ,0.25^\circ,0.5^\circ\}$ with the analytical MCRB and the CRB. Monte--Carlo MSE results are obtained over $4,000$ trials with a single snapshot per trial. For all non-zero 
spoofing offsets, the empirical MSE closely follows the proposed closed-form MCRB, demonstrating that the latter accurately bounds the achievable AoA estimation accuracy under 
spoofing. In contrast, the classical CRB fails to capture the SNR-independent error floor and continues to decrease with increasing SNR. In the absence of mismatch ($\Delta=0^\circ$), the MSE perfectly aligns with both the CRB and the MCRB across the entire SNR range, confirming MCRB’s efficiency under correct model specification and illustrating that a spatially collinear attacker does not introduce estimation penalty, making AoA-based discrimination infeasible at the aligned angle.

\begin{figure}[t]
    \centering
    \includegraphics[width=\linewidth]{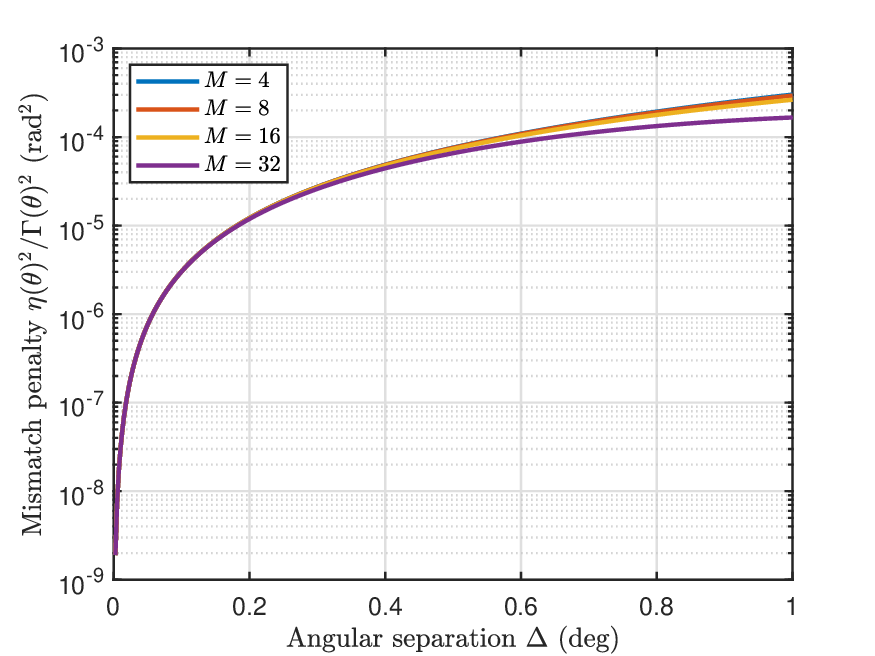}
    \caption{Mismatch of the MCRB versus angular offset
    $\Delta\in[0^{\circ},1^{\circ}]$.}
    \label{fig:penalty_vs_delta_M}
\end{figure}

Fig.~\ref{fig:penalty_vs_delta_M} illustrates the behavior of the mismatch penalty term of the MCRB, i.e., $\frac{\eta(\theta)^2}{\Gamma(\theta)^2}$ across different angular separations $\Delta\in[0^{\circ},1^{\circ}]$ between the legitimate transmitter and the spoofer, over $M=\{4,8,16,32\}$ antenna elements. As shown, for $\Delta>0$, the penalty term decreases with increasing number of antennas $M$, confirming that larger arrays mitigate the absolute impact of model mismatch. This behavior is consistent with the $\mathcal{O}(M^{-2})$ scaling of the mismatch term, in contrast to the $\mathcal{O}(M^{-3})$ decay of the classical CRB, indicating that at high SNR and/or large arrays the MCRB can be dominated by the mismatch-induced floor.

\begin{figure}[t]
    \centering
    \includegraphics[width=\linewidth]{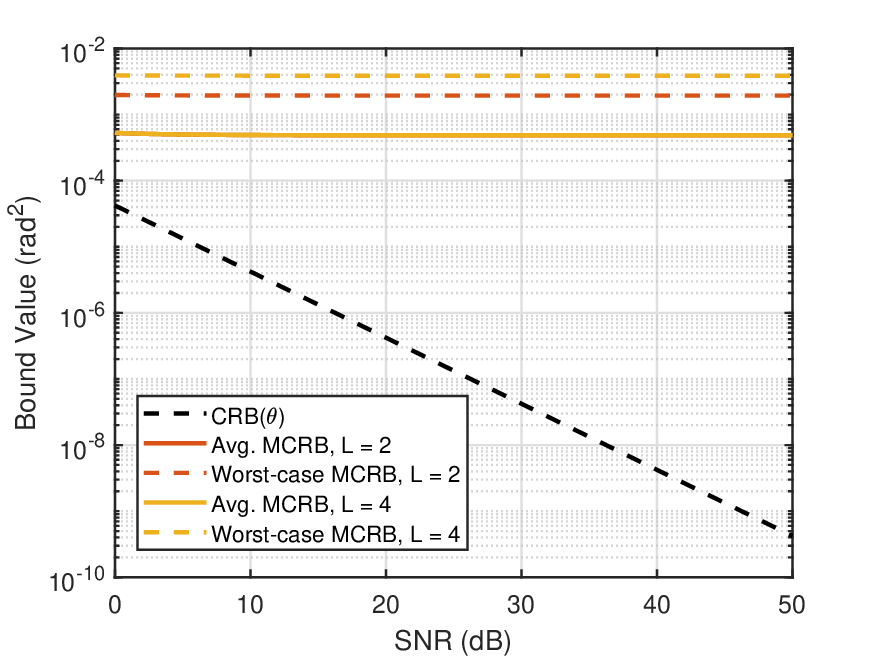}
    \caption{Average and worst-case MCRB versus SNR for AoA estimation under spoofing, with $M=16$ and $\Delta=0.5^{\circ}$.}
    \label{fig:L_SNR}
\end{figure}

Fig.~\ref{fig:L_SNR} illustrates the fundamental impact of an $L$-antenna spoofing adversary on AoA estimation accuracy for angular offset of $\Delta=0.5^{\circ}$, through the proposed MCRB. The adversary employs $L\in\{2,4\}$ antennas and generates the spoofed mean signal $\mathbf{s}=\sum_{\ell=1}^{L} q_{\ell}\mathbf{a}(\hat{\theta})$ under the power constraint $\sum_{\ell}|q_{\ell}|^{2}=1$. For the average MCRB, random precoding phases $q_{\ell}=\frac{1}{\sqrt{L}}e^{j\phi_{\ell}}$, $\phi_{\ell}\sim\mathcal{U}[0,2\pi)$, are used and results are averaged over multiple realizations. For the worst-case MCRB, the attacker phases are chosen to maximize the mismatch-induced term, yielding coherent accumulation of $\Im\{\sum_{\ell} q_{\ell} S_{M}(r_{\ell})\}$. As expected, the CRB decreases monotonically and vanishes at high SNR, while the MCRB exhibits the non-zero error floor that is independent of SNR. Moreover, the worst-case MCRB is significantly larger than the average MCRB for both $L=2$ and $L=4$, demonstrating that coordinated attacker precoding can severely degrade AoA estimation accuracy even for small angular offsets. Increasing the number of attacker antennas results in a moderate increase in the average error floor, but a more pronounced increase under worst-case precoding, highlighting the dominant role of phase alignment rather than antenna count alone.

\section{Conclusions}

We derived the MCRB for AoA estimation under spoofing starting from the likelihood of
the assumed Gaussian model for a ULA with \(M\) antennas and a single snapshot. The MCRB explicitly
exhibits a mismatch-induced penalty term that can create a non-vanishing error floor,
highlighting the fundamental impact of spoofing on AoA estimation accuracy.

\appendices
\section{Derivation of \(\eta(\theta)\)}

We now give an explicit analytic form of \(\eta(\theta)\) suitable for numerical
evaluation. From \eqref{eq:eta_def_repeat} and \eqref{eq:delta_def},
\begin{equation}
    \eta(\theta)
    = \Re\!\Big\{
    \sum_{\ell=1}^{L} q_{\ell}\,
    \dot{\mathbf a}(\theta)^{H}\mathbf a(\hat{\theta}_{\ell})
    - \dot{\mathbf a}(\theta)^{H}\mathbf a(\theta)
    \Big\}.
    \label{eq:eta_split}
\end{equation}
The self-term \(\dot{\mathbf a}(\theta)^{H}\mathbf a(\theta)\) is purely imaginary
(equal to \(j\kappa\cos\theta\cdot M(M-1)/2\)) and therefore does not contribute
to the real part. Hence,
\begin{equation}
    \eta(\theta)
    = \Re\!\Big\{
    \sum_{\ell=1}^{L} q_{\ell}\,
    \dot{\mathbf a}(\theta)^{H}\mathbf a(\hat{\theta}_{\ell})
    \Big\}.
    \label{eq:eta_no_self}
\end{equation}

We next compute the inner product
\(\dot{\mathbf a}(\theta)^{H}\mathbf a(\phi)\) for a general angle \(\phi\).
Using
\begin{equation}
    \dot{\mathbf a}(\theta)
    = -j\kappa\cos\theta\,
    \mathrm{diag}\{0,1,\ldots,M-1\}\,
    \mathbf a(\theta),
\end{equation}
we obtain
\begin{equation}
    \dot{\mathbf a}(\theta)^{H}\mathbf a(\phi)
    = j\kappa\cos\theta
    \sum_{m=0}^{M-1}
    m\,e^{j\kappa m(\sin\phi-\sin\theta)}.
\end{equation}
Defining the phase shift
\begin{equation}
    r \triangleq e^{j\kappa(\sin\phi-\sin\theta)},
\end{equation}
the finite weighted geometric sum (for \(r\neq 1\)) has the closed form
\begin{equation}
    \sum_{m=0}^{M-1} m r^{m}
    = \frac{r\big(1 - M r^{M-1} + (M-1) r^{M}\big)}{(1-r)^{2}}.
    \label{eq:geom_sum}
\end{equation}
If \(r=1\) (i.e., \(\phi=\theta\)), the sum equals \(M(M-1)/2\), which coincides
with the limit of \eqref{eq:geom_sum} as \(r\to 1\).

Applying the above with \(\phi=\hat{\theta}_{\ell}\), and defining
\(r_{\ell}\triangleq e^{j\kappa(\sin\hat{\theta}_{\ell}-\sin\theta)}\), yields
\begin{equation}
    \dot{\mathbf a}(\theta)^{H}\mathbf a(\hat{\theta}_{\ell})
    = j\kappa\cos\theta\,S_M(r_{\ell}),
    \qquad
    S_M(r_{\ell})\triangleq
    \sum_{m=0}^{M-1} m r_{\ell}^{m}.
\end{equation}
Hence, from \eqref{eq:eta_no_self} gives
\begin{equation}
    \eta(\theta)
    = \Re\!\Big\{
    j\kappa\cos\theta
    \sum_{\ell=1}^{L} q_{\ell}\,S_M(r_{\ell})
    \Big\}.
\end{equation}
Using \(\Re\{jz\}=-\Im\{z\}\) finally results in
\begin{equation}
    \eta(\theta)
    = -\kappa\cos\theta\;
    \Im\!\Big\{
    \sum_{\ell=1}^{L} q_{\ell}\,S_M(r_{\ell})
    \Big\},
    \label{eq:eta_final}
\end{equation}
where, for \(r_{\ell}\neq 1\),
\begin{equation}
    S_M(r_{\ell})
    = \frac{r_{\ell}\big(1 - M r_{\ell}^{M-1}
    + (M-1) r_{\ell}^{M}\big)}{(1-r_{\ell})^{2}},
\end{equation}
and for \(r_{\ell}=1\), \(S_M(1)=M(M-1)/2\) (use the limit).

\section{Acknowledgment}
A. Chorti has been partially supported by the EC through the Horizon Europe/JU SNS project ROBUST-6G (Grant Agreement no. 101139068), the IPAL Project CONNECTING, the TalCyb Chair in Cybersecurity and by the French government under the France 2030 ANR program “PEPR Networks of the Future” (ref. ANR-22-PEFT-0009). S. Skaperas has been supported by the Horizon Europe/JU SNS project ROBUST-6G (Grant Agreement no. 101139068).

\bibliographystyle{IEEEtran}
\bibliography{references}

@ARTICLE{Usman2025,
  author={Ali, Usman and Melazzi, Nicola Blefari and Bartoletti, Stefania},
  journal={IEEE Wireless Communications Letters}, 
  title={Cooperative ISAC Under Spoofing Attacks}, 
  year={2025},
  volume={14},
  number={9},
  pages={2683-2687},
  }

@INPROCEEDINGS{Pham_Globecom23,
  author={Pham, Thuy M. and Senigagliesi, Linda and Baldi, Marco and Fettweis, Gerhard P. and Chorti, Arsenia},
  booktitle={GLOBECOM 2023 - 2023 IEEE Global Communications Conference}, 
  title={Machine Learning-Based Robust Physical Layer Authentication Using Angle of Arrival Estimation}, 
  year={2023},
  volume={},
  number={},
  pages={13-18},
 }

@article{Pham_arxiv,
    author = {Thuy M. Pham and Linda Senigagliesi and Marco Baldi and Rafael F. Schaefer and Gerhard P. Fettweis and Arsenia Chorti},
    title = {Leveraging Angle of Arrival Estimation against Impersonation Attacks in Physical Layer Authentication},
    journal = arxiv,
    year = 2025
}

@article{abed2021misspecified,
  title={Misspecified Cramer--Rao bounds for blind channel estimation under channel order misspecification},
  author={Abed-Meraim, Karim and Trung, Nguyen Linh and others},
  journal={IEEE Transactions on Signal Processing},
  volume={69},
  pages={5372--5385},
  year={2021},
  publisher={IEEE}
}

@inproceedings{Zheng2023MCRB,
  author={Zheng, Pan and Ballal, Tarig and Chen, Hui and Al-Naffouri, Tareq Y.},
  booktitle={2023 IEEE International Conference on Acoustics, Speech and Signal Processing (ICASSP)}, 
  title={The Misspecified {Cram\'{e}r-Rao} Bound of {RIS}-Aided Localization Under Geometry Mismatch}, 
  year={2023},
  pages={1-5},
  doi={10.1109/ICASSP49357.2023.10095147}
}

@article{Varughese2025MCRB,
  author={Varughese, Maxine and Bartels, Randy and Pezeshki, Ali},
  journal={Optics Letters}, 
  title={Misspecified {Cram\'{e}r-Rao} lower bound with {Poisson} statistics and its application in localization microscopy}, 
  year={2025},
  volume={50},
  number={16},
  pages={5041-5044},
  doi={10.1364/OL.566313},
  month={Aug}
}

@article{fortunati2016misspecified,
  title={The misspecified Cram{\'e}r-Rao bound and its application to scatter matrix estimation in complex elliptically symmetric distributions},
  author={Fortunati, Stefano and Gini, Fulvio and Greco, Maria S},
  journal={IEEE Transactions on Signal Processing},
  volume={64},
  number={9},
  pages={2387--2399},
  year={2016},
  publisher={IEEE}
}

@article{hoang2024physical,
  title={Physical layer authentication and security design in the machine learning era},
  author={Hoang, Tiep M and Vahid, Alireza and Tuan, Hoang Duong and Hanzo, Lajos},
  journal={IEEE Communications Surveys \& Tutorials},
  volume={26},
  number={3},
  pages={1830--1860},
  year={2024},
  publisher={IEEE}
}

\end{document}